\documentclass[
aps,
prl,
twocolumn,
amsfonts,
amssymb,
amsmath,
letterpaper,
]{revtex4-1}
\usepackage{amsmath}
\usepackage{amssymb}
\usepackage%
{graphicx}
\usepackage{extdash}
\usepackage{epsfig}
\usepackage{textcomp}
\DeclareFontFamily{U}{euc}{}% I chose euc because the chart is called Euler cursive 
\DeclareFontShape{U}{euc}{m}{n}{<-6>eurm5<6-8>eurm7<8->eurm10}{}% 
\DeclareSymbolFont{AMSc}{U}{euc}{m}{n} % I chose AMSc because AMSa and AMSb are defined in the amsfonts-package 
\DeclareMathSymbol{\umu}{\mathord}{AMSc}{"16}

\newcommand{\ensuretext}[1]{\ensuremath{\text{#1}}}

\newcommand{\unit}[1]{\ensuretext{\textrm{\,}}\ensuremath{\mathrm{#1}}}

\newcommand{\Mum}{\ensuremath{\umu}\ensuremath{\mathrm{m}}}

\begin{document}

%\preprint{}

\title{Two surface plasmon decay of plasma oscillations}

\author{T.~Kluge}
\email[]{t.kluge@hzdr.de}
\homepage[]{http://hzdr.de/crp}
%\thanks{}
\affiliation{Helmholtz-Zentrum Dresden-Rossendorf, Germany}

\author{J.~Metzkes}
%\thanks{}
\affiliation{Helmholtz-Zentrum Dresden-Rossendorf, Germany}

\author{K.~Zeil}
\affiliation{Helmholtz-Zentrum Dresden-Rossendorf, Germany}

\author{M.~Bussmann}
\affiliation{Helmholtz-Zentrum Dresden-Rossendorf, Germany}

\author{U.~Schramm}
\affiliation{Helmholtz-Zentrum Dresden-Rossendorf, Germany}
\affiliation{TU Dresden, Germany}

\author{T.\,E.~Cowan}
\affiliation{Helmholtz-Zentrum Dresden-Rossendorf, Germany}
\affiliation{TU Dresden, Germany}

\date{\today}
\begin{abstract}
The interaction of ultra-intense lasers with solid foils can be used to accelerate ions to high energies well exceeding 60~MeV~\cite{Gaillard2011}. The non-linear relativistic motion of electrons in the intense laser radiation leads to their acceleration and later to the acceleration of ions. Ions can be accelerated from the front surface, the foil interior region, and the foil rear surface (TNSA, most widely used), or the foil may be accelerated as a whole if sufficiently thin (RPA). \\
Here, we focus on the most widely used mechanism for laser ion-acceleration of TNSA. Starting from perfectly flat foils we show by simulations how electron filamentation at or inside the solid leads to spatial modulations in the ions. The exact dynamics depend very sensitively on the chosen initial parameters which has a tremendous effect on electron dynamics. In the case of step-like density gradients we find evidence that suggests a two-surface-plasmon decay of plasma oscillations triggering a Raileigh-Taylor-like instability. 
\end{abstract}
\pacs{12345}
%\keywords{}

\maketitle
\begin{figure*}
  \centering
  \includegraphics[width=16cm]{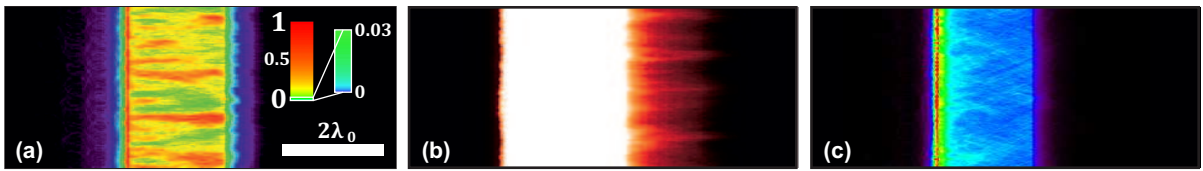}\\
  \caption{(a) Part of the electron energy density distribution at $t=18\pi=9\unit{T}_0$ after the laser , normalized to 2/3 of its maximum and (b) ion density distribution of simulation A, and (c) electron energy distribution of simulation B. Simulation parameters: 
only the plane defined by the laser propagation direction (x-axis) and the electric field polarization (y-axis) is simulated, assuming invariance of the system in z-direction; simulation box $10\lambda_0\times10\lambda_0$, 192 cells per $\lambda_0$ and laser period $\mathrm{T}_0$;  
laser is modeled by a transversely plane wave with short gaussian rise with 1 sigma-width of $2\lambda_0/\mathrm{c}$ followed by a flat top, wavelength $\lambda_0$, peak intensity $I_0\unit{[W/cm^2]}=1.38\times 10^{20}/(\lambda_0\unit{[\Mum]})^2$  corresponding to a normalized field strength $a_0=[2I_0/(\mathrm{n}_c\mathrm{m}_e\mathrm{c}^3)]^{1/2}=10$ where $\mathrm{n}_c=\mathrm{m}_e\epsilon_0\omega_0^2e^2$ is the critical electron density for the laser with angular frequency $\omega_0$;  %$a_0=[2eI_0\lambda_0^2/(4\epsilon_0\pi^2\mathrm{m}_e^2\mathrm{c}^5)]^{1/2}$
laser is aligned normal to the target foil surface;  
flat target foils modeled by a fully preionized plasma slab of $2\lambda_0$ thickness: ions with charge-to-mass ratio $Q/M=1/2$ and neutralizing electrons with density $\mathrm{n}_{e,0}=100\unit{n}_c$; 
front of the foils is modeled by an exponentially increasing preplasma with scale length $L$, \textbf{(a)} and \textbf{(b)} $L=2\pi/10=0.1\,\lambda_0$ (simulation A) and \textbf{(c)} $L=0$ (simulation B).}\label{fig:1}
\end{figure*}

In laser ion acceleration, an ultra intense laser interacts with a solid and quickly ionizes its surface. The laser light creates an immense pressure on the surface, produces large and fast electron currents into the solid and excites plasma waves. In such a scenario many instabilities develop which can disrupt the surface and break up the electron currents. \\
It has been shown previously that instabilities in \emph{ultrathin} solid foils can imprint onto the spatial structure of ions in the regime of relativistically induced transparency target normal sheath acceleration (RIT TNSA~\cite{TNSA}) or radiation pressure acceleration (RPA~\cite{Esirkepov-LaserPiston})~\cite{RT} -- mostly attributed to the Rayleigh-Taylor instability (RTI)~\cite{palmer}. Instabilities behind the foils can also develop during the ps timescales of the plasma expansion of TNSA~\cite{Quinn2012}. 
Here we show that instabilities developing in solid foils can also be of significance for the laser absorption and ion acceleration for thicker foils where no RIT occurs. 
We present results from particle-in-cell (PIC) simulations using the code i\textsc{picls2d}~\cite{picls} and refer the reader to our Ref.~\cite{Metzkes} for experimental evidence. 
We use dimensionless units with $c=e=\mathrm{m}_e=\omega_0=1$ and measure distances in $x$-direction (the laser direction) relative to the initial foil front surface position and times relative to the time when the laser intensity maximum reaches $x=0$. 
In most of our analysis we will focus on two simulations, comparing the cases of a flat foil with exponential preformed plasma on the surface (simulation A) and without (simulation B), the full simulation parameters are given in the caption of Fig.~\ref{fig:1}. 

\begin{figure*}
  \centering
  \includegraphics[width=\textwidth]{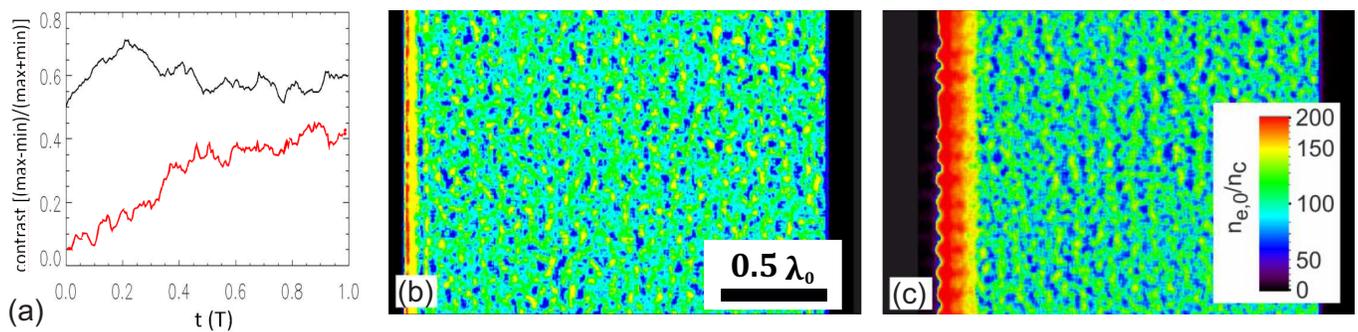}
  \caption{(a) Transverse electron energy density contrast when co-moving with one of the $2\omega_0$ electron slabs after it is injected into the solid for simulation A (black line) and simulation B (gray/red line). (b, c) Electron density snapshot at $t=18\pi=9\unit{T}_0$ for (b) simulation A and (c) simulation B. }\label{fig:3}
\end{figure*}

One main finding is that instabilities which are prone to evolve in all of our simulations imprint on the ion density distribution behind the foil slab even for the micron thick foils used. The instabilities we discuss in this paper are developing in the electrons first since the heavier ions remain almost at rest during the relevant time scales of electron motion. Of course this is not strictly correct since otherwise no RTI instability could develop. Fig.~\ref{fig:1}(a) shows exemplary part of the electron density distribution of simulation A $t=18\pi=9\unit{T}_0$ after the laser maximum has reached the foil front surface, and Fig.~\ref{fig:1}(b) shows the corresponding ion density distribution. In this simulation the electrons pushed into the foil at $2\omega_0$ by the laser Lorentz force are prone to a transverse Weibel-like instability~\cite{Sentoku2000}. The $2\omega_0$ electron slabs become filamented transversely and create energetic channels almost normal to the foil surface which are surrounded by strong quasi-static magnetic fields. This happens during the first few plasma wavelengths $\lambda_p\equiv 2\pi\omega_p^{-1}=2\pi\mathrm{n}_{e,0}^{-1/2}$ as can be seen in Fig.~\ref{fig:3}a. The ion acceleration at the foil rear surface then follows this structure(cp.~Fig,~\ref{fig:1}a,b). It is this that is different to the mechanism described in~\cite{Quinn2012}: the rear surface ion's acceleration itself is structured transversely \emph{from the beginning} since the driving energetic electrons reach the rear surface filamented already in contrast to the relatively slow Weibel instability witnessed in~\cite{Quinn2012} \emph{during} the ion acceleration. Of course both can happen concurrently. 

\begin{figure*}
  \centering
  \includegraphics[width=\textwidth]{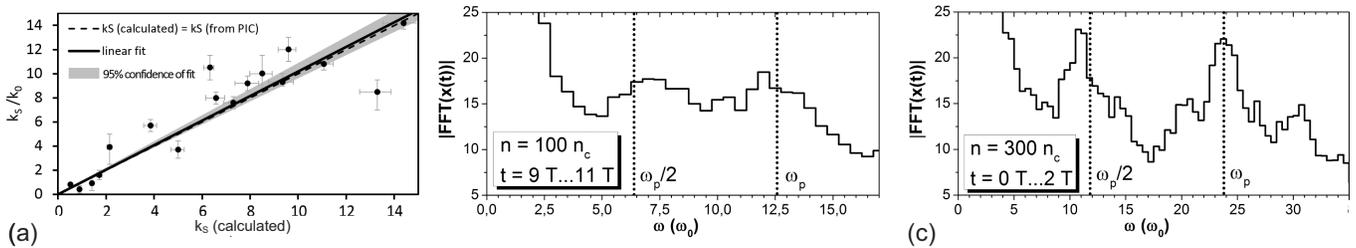}
  \caption{(a) $k_S^{PIC}$ versus calculated $k_S$ from [Eqn.~(6) in~\cite{Kaw1970}] not including thermal effects. (b) and (c) Temporal Fourier transform of surface oscillations for (b) simulation B and (c) a simulation with same parameters as B but $\mathrm{n}_{e,0}=300$ and $a_0=30$ (details see text). The time over which the oscillations were evaluated is indicted in the legend.}\label{fig:4}
\end{figure*}

\begin{figure}
  \centering
  \includegraphics[width=8cm]{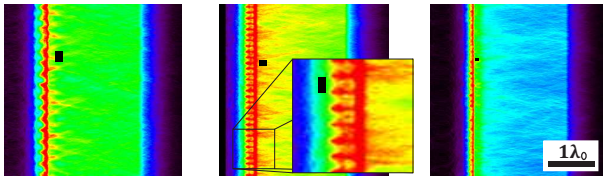}
  \caption{Energy density distribution for simulations with parameters as simulation B but with initial surface roughness. The largest spatial frequency is indicated by the black bar, the smallest is twice this size. The amplitude of the surface roughness is $0.1\lambda_0$. Color scale is the same as in Fig.~\ref{fig:1}. Largest surface ripple growth occurs when the spatial frequency $k_S$ is seeded, with a characteristic mushroom shape (non-linear RTI) growing at the front (central panel). Here, $k_S^{PIC}=7.6$, $k_S=7.3$.}\label{fig:5}
\end{figure}

We now study the case of step-like density gradient at the front surface (simulation B). 
In comparison to simulation A, where the bulk electrons remained distributed almost homogeneously transversely at the surface and only energetic electrons became filamented inside the foil, here the foil surface already shows transverse structure both in electron density (Fig.~\ref{fig:3}b,c) and energy density (Fig.~\ref{fig:3}a), with the latter not increasing during passage through the foil contrarily to simulation A. The reason for this lies in the density ripples which give rise to filamented injection of electrons, similar to the mechanism described in~\cite{Bigongiari2011}. 
We therefore conclude that in simulation B another filamentation mechanism must be present.
Density ripples usually are created by RTI and the density distribution in Fig.~\ref{fig:3}c shows a pattern consistent with the linear stage of RTI. 
Yet, typically the average distance $\lambda_S$ between density maxima (wave number $k_S=\lambda_S^{-1}$) should be given by the smallest transverse scale that could seed the instability, since the growth rate $\Gamma\cong(gk_S)^{-1/2}$ ($g$ is the acceleration of the surface) is largest for the largest $k_S$. 
Hence we are looking for a mechanism that can seed a transverse density fluctuation which is then prone to a RTI growth~\cite{Zhang2011,Macchi2015}. 
Such a mechanism was described e.g. in~\cite{Macchi2001,Macchi2002} where a generation of 2D electron surface oscillations (2DESOs) by a parametric decay of $2\omega_0$ oscillations of the foil surface in the laser field was found. 
There from initially purely longitudinal driving oscillations of the surface (for the $2\omega_0$ oscillations the driver frequency is $\omega_D=2\omega_0$), oscillations in a standing wave along the surface are generated via excitation of upward and downward moving and interfering surface waves with $\omega_{S}=\omega_D/2$ and $k_{S}$ given by [Eqn.~(6) in~\cite{Kaw1970}]. 
%Maxima of the standing wave oscillation amplitude are then separated by $k_S=2k_{SW}$. 
For our simulation conditions this would correspond to $k_S\cong 1$, much smaller than the value extracted from the PIC simulation $k_S^{PIC}\cong 7.7$. This decay of $2\omega_0$ oscillations with the characteristic $1\omega_0$ longitudinal oscillation is observed in our simulation only at later times growing only slowly, which we attribute to larger $k_S$. 
However, considering not the $2\omega_0$ oscillation as a driving source but the plasma oscillations excited by the disturbances we find good numerical agreement. Fig.~\ref{fig:4} shows a comparison of $k_S^{PIC}$ for a number of simulations we performed with different parameters (varying $\mathrm{n}_{e,0}$ and $a_0$) with the analytical value $k_S$. 
The agreement is generally rather good. 
%  \caption{(a) $k_S^{PIC}$ versus calculated $k_S=2k_{SW}$ with $k_{SW}$ from [Eqn.~(6) in~\cite{Kaw1970}] including thermal effects. (b) and (c) Temporal Fourier transform of surface oscillations for (b) simulation B and (c) a simulation with same parameters as B but $\mathrm{n}_{e,0}=300$ and $a_0=30$ (details see text). The time over which the oscillations were evaluated is indicted in the legend.}\label{fig:4}
Also, we find oscillations at $\omega_p/2$ in the spectrum of surface oscillations, Fig.~\ref{fig:4}(b). Here we defined a density level and recorded the position in x-direction (laser direction) where the rising density reaches that value. We then computed the temporal Fourier transform of this position averaged over the positions along the y-direction at the density ripple maxima, neglecting the regions in between. 
The result would be a strong signal around the laser harmonics (overlying all other signals) which however are expected to be transversely flat here. 
Therefore we took the oscillation at the "`inner surface"'.   
By this we mean the oscillation of the falling density slope behind the region of laser pressure steepened electron density, which can screen the lower laser harmonics.  
As expected we find a signal at the plasma frequency $\omega_p=(\mathrm{n}_{e,max}/T_{max})^{1/2}$ computed using the maximum density and temperature at the peak of the laser pressure steepened density profile averaged over the distance of a plasma period. 
Additionally we also found another pronounced peak at $\omega_p/2$. 
Since there is still a significant contribution from laser harmonics, we performed another simulation with $\mathrm{n}_{e,0}=300$ and $a_0=30$ (Fig.~\ref{fig:4}(c)). 
Due to the higher density the skin depth $\delta\cong c \omega_p^{-1}$ is much shorter, screening laser harmonics, and additionally the plasma frequency and its half value are shifted to higher frequencies and hence are at higher and hence weaker harmonics. 
Here we again obtain a very clear signal at $\omega_p$ and $\omega_p/2$. \\
It is interesting to mention that the development of surface ripples seen in simulation B can be artificially seeded. 
Introducing an initial surface roughness containing a mixture of several spatial frequencies around $k_S=2k_{SW}$ we first find that the smallest spatial frequencies have the fastest growth, as expected for RTI, and that this growth is most dominant when the spatial frequencies are initialized around $k_S$~(Fig.~\ref{fig:5}).  
For yet higher spatial frequencies the effect of seeding abruptly ceases. \\
Summarizing, we found by simulations that the laser ion acceleration process in the TNSA regime can be influenced by electron instabilities at the surface or inside the foil. 
The exact mechanism and scale of spatial modulations in the beam of hot electrons and ions sensitively depends on the initial parameters. 
For the pursuit of higher ion energies at higher laser intensities, e.g. available at
(future) Petawatt laser systems, a deeper understanding and further characterization
of laser and target parameter dependent plasma instabilities will be essential. 

The work has been partially supported by EC FP7 LASERLAB-EUROPE/CHARPAC (contract
284464) and by the German Federal Ministry of Education and Research (BMBF) under
contract number 03Z1O511. 

\section*{References}
\providecommand{\newblock}{}

\end{document}